\documentclass{Interspeech}

\interspeechcameraready

\usepackage{comment}
\usepackage{enumitem,amssymb}
\usepackage[utf8]{inputenc} %
\usepackage[T1]{fontenc}    %
\usepackage{hyperref}       %
\usepackage{url}            %
\usepackage{booktabs}       %
\usepackage{amsfonts}       %
\usepackage{nicefrac}       %
\usepackage{microtype}      %
\usepackage[table]{xcolor}  %
\usepackage{soul}
\usepackage{cite}

\usepackage{arydshln}
\usepackage{enumitem}
\usepackage{amsmath}
\usepackage{setspace} %
\usepackage{multirow} %
\usepackage[labelfont=bf]{caption} %
\usepackage{graphicx} %
\usepackage{scalerel} %

\renewenvironment{thebibliography}[1]{
  \begin{oldthebibliography}{#1}
  \setlength{\itemsep}{-0.05em}
  \setlength{\parskip}{0.em}
}
{
  \end{oldthebibliography}
}

\makeatletter
\def\bstctlcite{\@ifnextchar[{\@bstctlcite}{\@bstctlcite[@auxout]}}
\def\@bstctlcite[#1]#2{\@bsphack
  \@for\@citeb:=#2\do{%
    \edef\@citeb{\expandafter\@firstofone\@citeb}%
    \if@filesw\immediate\write\csname #1\endcsname{\string\citation{\@citeb}}\fi}%
  \@esphack}
\makeatother

\title{Interspeech 2025 URGENT Speech Enhancement Challenge}

\author[affiliation={1}]{Kohei}{Saijo}
\author[affiliation={2}]{Wangyou}{Zhang}
\author[affiliation={3}]{Samuele}{Cornell}
\author[affiliation={4}]{Robin}{Scheibler}
\author[affiliation={2}]{Chenda}{Li}
\author[affiliation={5}]{Zhaoheng}{Ni}
\author[affiliation={5}]{Anurag}{Kumar}
\author[affiliation={6}]{Marvin}{Sach}
\author[affiliation={6}]{Yihui}{Fu}
\author[affiliation={2}]{Wei}{Wang}
\author[affiliation={6}]{Tim}{Fingscheidt}
\author[affiliation={3}]{Shinji}{Watanabe}

\affiliation{}{%
  Waseda University, Japan
  $^2$Shanghai Jiao Tong University, China
  $^3$Carnegie Mellon University, USA
  $^4$Google DeepMind, Japan
  $^5$Meta, USA
  $^6$Technische Universität Braunschweig}{Germany}
\email{saijo@pcl.cs.waseda.ac.jp}

\newlist{todolist}{itemize}{2}
\setlist[todolist]{label=$\square$}

\begin{document}

\maketitle

\begin{abstract}

There has been a growing effort to develop universal speech enhancement (SE) to handle inputs with various speech distortions and recording conditions. The URGENT Challenge series aims to foster such universal SE by embracing a broad range of distortion types, increasing data diversity, and incorporating extensive evaluation metrics. This work introduces the Interspeech 2025 URGENT Challenge, the second edition of the series, to explore several aspects that have received limited attention so far: language dependency, universality for more distortion types, data scalability, and the effectiveness of using noisy training data. We received 32 submissions, where the best system uses a discriminative model, while most other competitive ones are hybrid methods. Analysis reveals some key findings: (i) some generative or hybrid approaches are preferred in subjective evaluations over the top discriminative model, and (ii) purely generative SE models can exhibit language dependency.

\end{abstract}

\bstctlcite{IEEEexample:BSTcontrol}

\section{Introduction}

Speech enhancement (SE) aims to improve speech quality degraded by noise, reverberation, or other distortions.
It has seen significant progress thanks to neural networks.
However, most prior work has focused on matched training–inference conditions and limited tasks such as noise suppression and dereverberation.

In recent years, the advent of generative models has broadened the scope of SE to include more complex tasks such as bandwidth extension, packet loss concealment~\cite{INTERSPEECH2022-Diener2022}, and wind noise reduction~\cite{lemercier2023wind}. 
Following this trend, there has been growing interest in developing universal SE models capable of handling multiple tasks and diverse input formats within a single framework.
For example, score-based diffusion models have been explored for universal SE~\cite{Universal-Serra2022, scheibler24_interspeech}, enabling a single model to address multiple SE tasks.
To support diverse input formats, models that process multiple sampling rates and varying numbers of microphones have also been proposed~\cite{Toward-Zhang2023}.
However, despite these advances, a well-established benchmark for evaluating generative and universal SE models remains lacking, making fair comparisons challenging.

To address this gap, the URGENT Challenge (Universality, Robustness, and Generalizability of speech EnhancemeNT) was launched.
The first edition, the URGENT 2024 Challenge~\cite{URGENT-Zhang2024}, aimed to lay the foundation for developing universal SE models by setting two key objectives: (i) handling four types of distortion (additive noise, reverberation, bandwidth limitation, and clipping) and (ii) accommodating input signals with varying sampling rates.
The challenge utilized a collection of publicly available datasets, significantly larger than those used in most prior studies, and employed a comprehensive evaluation framework with as many as 13 evaluation metrics.

To further advance the development of universal SE models, we propose the second edition of the series, Interspeech 2025 URGENT Challenge.
This iteration introduces the following modifications based on our preliminary investigations~\cite{lessons_learned_URGENT2024}:
\begin{itemize}
    \item \textbf{More data diversity}: There is room for improvement in robustness on, e.g., samples with unseen or rarely seen noise. To address this, the proposed challenge by incorporating more diverse speech and noise data.

    \item \textbf{Leveraging noisy data}: While noisy data is crucial for scaling up the amount of data due to the scarcity of clean data, we found that simple filtering methods (e.g., using DNSMOS scores~\cite{DNSMOS-Reddy2022}) did not remove noisy samples very effectively. We intentionally include more noisy data to encourage participants to explore effective ways of utilizing it.

    \item \textbf{Two tracks with different training data scales}: To examine the impact of data scale, the proposed challenge introduces two tracks with different training data sizes: one with $\sim$2.5k hours and the other with $\sim$60k hours of speech..
    
    \item \textbf{More distortions}: We found that SE models struggle to generalize to unseen distortions. To enhance generalizability, we consider three additional distortions, commonly observed in real recordings due to the recording device or environment.

    \item \textbf{Multilingual data}: While the first challenge used only English data, the proposed challenge incorporates multilingual data, allowing us to examine the accessibility of SE models and evaluation metrics across the languages.

\end{itemize}

The challenge has received 22 submissions for one track and 10 for the other.
This paper details the challenge design and presents a preliminary analysis of the submissions.
The details of the challenge are also available on our website\footnote{\label{fn:website_link}\url{https://urgent-challenge.github.io/urgent2025/}}.

\section{Related works}

Existing SE challenges have greatly advanced SE studies for specific scenarios, such as denoising and dereverberation~\cite{INTERSPEECH2020-Reddy2020,ICASSP-Dubey2023}, speech restoration~\cite{ICASSP-Cutler2023,ICASSP-Ristea2024}, packet loss concealment~\cite{INTERSPEECH2022-Diener2022}, and so on.
The URGENT challenge complements existing ones by focusing on universality, generalizability, and robustness across diverse scenarios and evaluation metrics.

Most previous SE studies use only monolingual data, and thus language dependency of SE models is still under-explored.
While the language dependency of discriminative SE has been reported to be small~\cite{wang22c_interspeech}, that of generative SE lacks thorough investigation.
Additionally, although often overlooked, some metrics can be language dependent~\cite{chong2005methodology,konane2021impact}.
Through this challenge, we collect the objective and subjective evaluation scores of various SE systems in multiple languages, which will later be analyzed to assess the language dependency of metrics\footnote{\label{fn:mos_footnote}We were able to collect the subjective listening scores only on English data during the challenge period due to the Interspeech challenge time constraint. We plan to collect data for other languages and analyze the metrics after the challenge concludes.}.

One of the focuses of this challenge is how to leverage possibly noisy data.
While there are some possible approaches such as data filtering based on non-intrusive SE metrics~\cite{DNSMOS-Reddy2022, NISQA-Mittag2021, UTMOS} or unsupervised learning~\cite{fujimura2021noisy,mixit}, a comprehensive comparison of them is currently lacking, particularly on large-scale data.
The proposed challenge aims to clarify which of these methods are more effective, or if new ones can be devised, by encouraging participants to explore how to leverage noisy data.

This challenge also explores the impact of training data size on final performance.
While this has been investigated in other fields~\cite{kaplan2020scaling, zhai2022scaling}, to our best knowledge, only an investigation on small-scale data has been done in SE field~\cite{zhang24i_interspeech}.
The proposed challenge offers two tracks with different data amounts, encouraging the investigation of data scalability.

\vspace{-1mm}
\section{Challenge Design}

\vspace{-1mm}
\subsection{Task definition}
\vspace{-1mm}
In the URGENT challenge, the SE process is defined as~\cite{URGENT-Zhang2024}:
{\setlength{\abovedisplayskip}{3pt}
\setlength{\belowdisplayskip}{3pt}
\begin{align}
	\hat{\mathbf{x}} &= \operatorname{SE}(\mathcal{F}(\mathbf{x})) \,, \label{eq:task}
\end{align}
}where $\mathbf{x}$ and $\hat{\mathbf{x}}$ are the desired and enhanced speech signals.
$\mathcal{F}(\cdot)$ is the distortion model that degrades the desired signal.
Our definition differs from the commonly adopted one in the literature in that the SE model has to handle
(i) inputs with \emph{various sampling frequencies (SF)} (8, 16, 22.05, 24, 32, 44.1, and 48kHz), while most existing SE systems often only consider a fixed SF,
(ii) inputs with \emph{seven types of distortions} (detailed in Section~\ref{ssec: distortions}), while most existing SE systems consider only noise or reverberation, and
(iii) \textit{multilingual speech}, while most existing SE studies consider only monolingual data.

\vspace{-1.5mm}
\subsection{Distortions}
\label{ssec: distortions}
\vspace{-1mm}

The challenge considers the following seven distortions in the distortion model $\mathcal{F}$: additive noise, reverberation, clipping, bandwidth limitation, codec loss (MP3 and OGG), packet loss, and wind noise\footnote{We used the wind-noise simulator provided in~\cite{lemercier2023wind}: \url{https://github.com/sp-uhh/storm}, which is based on \url{https://github.com/audiolabs/SC-Wind-Noise-Generator}}.
Compared to the first challenge, which included only the first four distortions~\cite{URGENT-Zhang2024}, the present challenge introduces three more challenging but realistic conditions.
Each noisy speech is degraded by up to five distortions simultaneously.

\vspace{-1.5mm}
\subsection{Data}
\label{ssec: data}
\begin{table}[t]
   \setstretch{0.86}
    \caption{
        Corpora used for training/validation/non-blind testing, where shaded cells are used only for non-blind testing\protect\footnotemark. 
        Data amounts of MLS and CommonVoice are limited in the first track but not in the second.
        $\lozenge$: We crawled the high-quality (less compressed) version of MLS (MLS-HQ) from LibriVox.
    }
    \vspace{-3.5mm}
    \label{tab: corpora}
    \centering
    \setlength{\tabcolsep}{2pt}
    \resizebox{1.0\columnwidth}{!}{%
        \begin{tabular}{l|lc} 
        \toprule
        {\textbf{Type}} & {\textbf{Corpus}} & {\textbf{Duration}} \\
        
        \midrule
        
        \multirow{8}{*}{Speech}& LibriVox data from DNS5 Challenge (en)~\cite{ICASSP-Dubey2023} & \textasciitilde{}350 h \\
        
        & LibriTTS reading speech (en)~\cite{LibriTTS-Zen2019} & \textasciitilde{}200 h \\
        
        & VCTK reading speech (en)~\cite{VCTK-Veaux2013} & \textasciitilde{}80 h\\
        
        & WSJ reading speech (en)~\cite{WSJ0-LDC1993,WSJ1-Consortium1994} & \textasciitilde{}85 h \\
        
        & EARS expressive speech (en)~\cite{richter2024ears} &\textasciitilde{}107 h \\
        
        & MLS-HQ$^\lozenge$ (en, de, fr, es)~\cite{pratap2020mls}  &\textasciitilde{}450 (48600) h \\

        & CommonVoice 19.0 (en, de, fr, es, zn) ~\cite{CommonVoice-Ardila2020} &\textasciitilde{}1300 (9500) h \\

        \midrule
        
        \multirow{6}{*}{Noise}& Noise from DNS5 Challenge~\cite{ICASSP-Dubey2023} & \textasciitilde{}180 h \\
        
        & WHAM! noise~\cite{WHAM-Wichern2019} & \textasciitilde{}70 h \\
        
        & FSD50K~\cite{fonseca2021fsd50k} &\textasciitilde{}100 h  \\
        
        & Free Music Archive (FMA, Medium partition)~\cite{defferrard2016fma} &\textasciitilde{}200 h \\
        
        & Wind noise simulated by participants  & -  \\
        
        \hdashline
        
        \rowcolor[HTML]{EEEEEE}& DEMAND~\cite{thiemann2013diverse} & - \\

        \rowcolor[HTML]{EEEEEE}& TUT Urban Acoustic Scenes 2018~\cite{mesaros2018multi} & - \\
        
        \midrule
        
        \multirow{3}{*}{RIR} & Simulated RIRs from DNS5 Challenge & \textasciitilde{}60k samples \\
        & Other RIRs simulated by participants & - \\
        
        \hdashline
       
        \rowcolor[HTML]{EEEEEE}& SLR28 real RIRs & -  \\
        \rowcolor[HTML]{EEEEEE}& BRUDEX~\cite{fejgin2023brudex} & - \\
        \rowcolor[HTML]{EEEEEE}& MYRiAD V2 econ~\cite{dietzen2023myriad} & - \\
       
        \bottomrule
        \end{tabular}%
    }
    \vspace{-6mm}
\end{table}
\footnotetext{This research did not train on any music, except to the extent that some music data from the FSD50K and FMA datasets was processed as background noise for speech enhancement purpose.}

\vspace{-1mm}
\subsubsection{Training data}
\label{sssec: train_data}
\vspace{-1mm}
A collection of public corpora listed in Table~\ref{tab: corpora} is used as training data.
Participants can freely simulate noisy speech using these data.
It is also allowed to freely simulate distortions mentioned in Section~\ref{ssec: distortions}.
However, to allow a fair comparison, using datasets other than those listed in Table~\ref{tab: corpora} is prohibited.

As some speech corpora include noisy speech, we preprocessed them following the methodology of the first challenge~\cite{URGENT-Zhang2024}.
Specifically, for DNS5 Challenge data and CommonVoice, we (i) resampled the speech to the lowest SF that can fully cover the effective frequency range, (ii) filtered out non-speech or samples dominated by silence using voice activity detection (VAD) \footnote{\url{https://github.com/wiseman/py-webrtcvad}}, and (iii) filtered out noisy samples based on the DNSMOS score~\cite{DNSMOS-Reddy2022}.
We also applied the resampling process to LibriTTS, MLS-HQ, DNS noise, FSD50K, and FMA.

Note that the VAD- and DNSMOS-based data filtering mentioned above is imperfect. 
Indeed, we found many samples with audible background noise still remaining after the data filtering.
Although such noisy samples may adversely affect SE model training, we intentionally keep such samples to encourage participants to explore how to leverage noisy samples effectively.

To investigate the language dependency of SE models, we use two multilingual corpora, MLS-HQ\footnote{We prepared the high-quality version of MLS (MLS-HQ) by crawling less-compressed speech from LibriVox (\url{https://librivox.org}) and segmenting them using the same timestamps as MLS.} and CommonVoice\,19.0 which cover five languages in total (en, de, fr, es, and zn).

\vspace{-1mm}
\subsubsection{Validation/non-blind test data}
\label{sssec: val_and_nonblind_test_data}
\vspace{-1mm}
In the challenge, it is allowed to freely simulate participants' own validation sets using the validation portion of the corpora in Table~\ref{tab: corpora} to select the system to submit.
However, we provide the official validation set containing 1000 noisy speech to allow a fair comparison.
When making the official validation set, since many CommonVoice samples are noisy, we manually selected only clean ones.
Samples from the other datasets are randomly selected after applying the data filtering mentioned in Section~\ref{sssec: train_data}.
The non-blind test set is made in a similar manner as the official validation set using the test portion of the corpora in Table~\ref{tab: corpora}.
However, we add some noise and RIRs shown in the shaded cells of Table~\ref{tab: corpora} to gauge robustness against unseen data.

\vspace{-1mm}
\subsubsection{Blind test data}
\label{sssec: blind_data}
\vspace{-1mm}

The blind test set, used to determine the final ranking, is primarily from domains unseen during training to assess the robustness and generalizability.
Specifically, the blind test set
\begin{itemize}
    \item Consists of 50\% simulated data and 50\% real recordings, both sourced from publicly available datasets. All synthetic data and most real recordings come from datasets not listed in Table~\ref{tab: corpora}. Since real recordings with certain distortions (e.g., packet loss) are scarce, we artificially apply clipping, packet loss, bandwidth limitations, and codec loss.
    \item Includes Japanese data as an unseen language.
    \item Has 150 samples per language (totaling 900 samples).
    \item Considers also unseen distortions due to neural audio codecs, specifically Encodec~\cite{defossez2022encodec} and Descript Audio Codec~\cite{kumar2023dac}, as codec lossy distortion in addition to MP3 and OGG.
\end{itemize}

\subsection{Two tracks with different data scales}
We provide two tracks with different training data scales but the same test set:
\begin{itemize}
    \item \texttt{Track 1}: We limit the duration of MLS and CommonVoice, resulting in $\sim$2.5k hours of speech.
    \item \texttt{Track 2}: We do not limit the duration of MLS and CommonVoice datasets, resulting in $\sim$60k hours of speech.
\end{itemize} 
This design allows us to explore (i)~the impact of data scale and (ii)~data-hungry methods such as large-scale self-supervised-learning-based methods.

\subsection{Evaluation metrics and ranking strategy}
We use multiple evaluation metrics to perform a more comprehensive evaluation.
For instance, the hallucination of a generative model could be hard to detect by non-intrusive metrics but is easily penalized by other metrics\footnote{Here, "hallucination" refers to discrepancies in spoken content or speaker characteristics between the noisy speech and enhanced speech.}. 
We adopt the following 14 metrics categorized into five categories:
\begin{itemize}[left=3.5pt]
    \item[1)] \textbf{Non-intrusive SE metrics}: DNSMOS~\cite{DNSMOS-Reddy2022}, NISQA~\cite{NISQA-Mittag2021}, and UTMOS~\cite{UTMOS}.
    \item[2)] \textbf{Intrusive SE metrics}: perceptual objective listening quality assessment (POLQA)~\cite{POLQA-Beerends2013}, perceptual evaluation of speech quality (PESQ)~\cite{PESQ-Rix2001}, extended short-time objective intelligibility (ESTOI)~\cite{ESTOI-Jensen2016}, signal-to-distortion ratio (SDR)~\cite{SDR-Vincent2006}, mel cepstral distortion (MCD)~\cite{MCD-Kubichek1993}, and log-spectral distance (LSD)~\cite{LSD-Gray1976}.
    \item[3)] \textbf{Downstream-task-independent metrics}: Levenshtein phone similarity (LPS)~\cite{LPS} and SpeechBERTScore (SBS)~\cite{SpeechBERTScore-Saeki2024} with mHuBERT-147~\cite{boito2024mhubert}.
    \item[4)]\textbf{Downstream-task-dependent metrics}: speaker similarity (SpkSim) with RawNet3~\cite{Pushing-Jung2022} and character accuracy (CAcc, 1 - character error rate) with OWSM v3.1~\cite{OWSMv3.1-Peng2024}.
    \item[5)] \textbf{Subjective metric}: absolute category rating mean opinion score (MOS) via ITU-T P.808 test~\cite{P.808,Open-Naderi2020}, implemented on the Amazon Mechanical Turk (on only English data\textsuperscript{\ref{fn:mos_footnote}}). %
\end{itemize}
Compared to the first challenge~\cite{URGENT-Zhang2024}, we made the following changes to support multilingual data: (i) we use an SSL model trained on multilingual data, mHuBERT-147, in SBS, and (ii) we evaluate CAcc instead of word accuracy.
In addition, UTMOS is included, as it showed a high correlation with MOS scores in our preliminary analysis~\cite{lessons_learned_URGENT2024}.
POLQA and MOS were evaluated only in the blind-testing phase.
The final ranking is obtained by aggregating the scores of the 14 metrics mentioned above with the following three steps inspired by the Friedman test~\cite{biUse-Friedman1937}:
\begin{itemize}[left=5pt]
    \item[1.] Calculate the ranking for each metric.
    \item[2.] Average the rankings of the metrics in each category.
    \item[3.] Average the category-wise rankings obtained in Step 2.
\end{itemize}
The overall ranking scores obtained in Step 3 are used to determine the final ranking.

\subsection{Baseline system}
We train the TF-GridNet model~\cite{TF_GridNet-Wang2023} with around 8.5M parameters on \texttt{Track1} training data and provide it as the baseline system.
We initialize the model with the pre-trained weights provided in the first challenge to save training time.
Training takes around 2.5 days using a single NVIDIA RTX A6000 GPU.
The training pipeline is based on ESPnet-SE library~\cite{ESPnet_SE-Li2021}. Please refer to the URGENT challenge website\textsuperscript{\ref{fn:website_link}} for more details.

\section{Results}
\label{sec: results}

\begin{table*}[t]
\setlength{\tabcolsep}{2.5pt}
\caption{
    Results of Track1 (\texttt{T*}) and Track2 (\texttt{T*}'), where the same numbers in IDs indicate systems submitted by the same team.
    Only 12 out of 22 submissions to Track1 and 4 out of 10 submissions to Track2 are shown due to space limitation.
    $\dagger$ in Track2 denotes the same submission as Track1.
    \texttt{T10} and \texttt{T22} are baseline and noisy input, respectively. 
    D, G, and D+G in `Model type` cell denote discriminative, generative, and hybrid of them, respectively.
    Numbers next to metric scores are ranking in each metric in each track.
    Note that MOS evaluation is done only for English data.
}
\vspace{-3.5mm}
\label{tab: results_track1and2}
\centering
\resizebox{\linewidth}{!}{

\begin{tabular}{lll|lll|llllll|ll|ll|l|l}
\toprule

\textbf{Rank} & \textbf{ID} &\textbf{Model} &\multicolumn{3}{c|}{\textbf{Non-intrusive SE metrics}} &\multicolumn{6}{c|}{\textbf{Intrusive SE metrics}} &\multicolumn{2}{c|}{\textbf{Down.-task-indep.}} &\multicolumn{2}{c|}{\textbf{Down.-task-dep.}} &\textbf{Subject.} &\textbf{Ranking}  \\

   & &\textbf{type} &\textbf{DNSMOS}$^\uparrow$ &\textbf{NISQA}$^\uparrow$ &\textbf{UTMOS}$^\uparrow$ &\textbf{POLQA}$^\uparrow$ &\textbf{PESQ}$^\uparrow$ &\textbf{ESTOI}$^\uparrow$ &\textbf{SDR}$^\uparrow$ &\textbf{MCD}$^\downarrow$ &\textbf{LSD}$^\downarrow$ &\textbf{SBS}$^\uparrow$ &\textbf{LPS}$^\uparrow$ &\textbf{SpkSim}$^\uparrow$ &\textbf{CAcc (\%)}$^\uparrow$ &\textbf{MOS}$^\uparrow$ &\textbf{score}$^\downarrow$ \\

\midrule

  1 &\texttt{T1} &D &  2.88 (8) &3.22 (6) &2.09 (5) &\textbf{3.40 (1)} &\textbf{2.64 (1)} &\textbf{0.82 (1)} &\textbf{12.66 (1)} &3.67 (2) &2.93 (3) &\textbf{0.87 (1)} &\textbf{0.74 (1)} &\textbf{0.76 (1)} &\textbf{79.80 (1)} &3.24 (5) &2.97 (1) \\
  2 &\texttt{T2} &D+G &  2.92 (5) &3.24 (5) &2.16 (3) &3.17 (4) &2.47 (4) &0.79 (5) &11.10 (5) &3.96 (7) &2.99 (8) & 0.84 (4) &0.70 (5) &0.74 (3) &76.06 (6) &3.32 (3) &4.37 (2) \\
  3 &\texttt{T3} &D+G &  2.94 (4) &3.25 (4) &2.19 (2) &3.16 (5) &2.45 (6) &0.79 (4) &11.25 (4) &4.79 (11) &3.66 (12) & 0.83 (6) &0.71 (3) &0.71 (6) &77.09 (5) &3.44 (2) &4.47 (3) \\
  4 &\texttt{T4} &D+G & 2.80 (15) &3.01 (10) &2.04 (6) &3.22 (2) &2.47 (3) &0.80 (3) &11.47 (3) &3.90 (5) &2.94 (4) & 0.85 (2) &0.71 (4) &0.74 (2) &78.06 (3) &3.28 (4) &4.63 (4) \\
  5 &\texttt{T5} &D+G & 2.83 (13) &2.92 (12) &2.03 (7) &3.18 (3) &2.48 (2) &0.81 (2) &11.69 (2) &\textbf{3.64 (1)} &2.98 (7) & 0.85 (3) &0.72 (2) &0.74 (4) &78.35 (2) &3.04 (10) &5.80 (5) \\
  6 &\texttt{T6} &D+G &  2.91 (7) &3.28 (3) &1.98 (9) &2.99 (8) &2.29 (8) &0.78 (6) &10.58 (7) &4.22 (8) &3.78 (13) & 0.83 (7) &0.70 (6) &0.70 (7) &77.15 (4) &3.21 (6) &6.53 (6) \\
  7 &\texttt{T7} &D+G & 2.85 (11) &3.08 (8) &1.98 (8) &3.13 (6) &2.45 (5) &0.78 (7) &10.74 (6) &3.95 (6) &2.81 (2) & 0.84 (5) &0.68 (7) &0.70 (9) &75.28 (9) &3.19 (7) &7.27 (7) \\
  8 &\texttt{T8} &D &  2.92 (6) &3.16 (7) &1.97 (10) &2.99 (7) &2.25 (9) &0.76 (9) &10.39 (8) &3.89 (4) &2.95 (5) & 0.82 (8) &0.66 (9) &  0.69 (10) &75.30 (8) &3.17 (9) &8.23 (8) \\
  9 &\texttt{T9} &D+G &  3.08 (2) &3.69 (2) &2.11 (4) &2.67 (14) &1.99 (14) &0.74 (12) &\hspace*{1.5 mm }7.43 (16) &4.42 (9) &3.14 (9) &0.80 (11) &  0.65 (10) &0.71 (5) &72.47 (15) &3.17 (8) &8.70 (9) \\
\rowcolor[HTML]{EEEEEE} 10 &\texttt{T10} &D & 2.85 (10) &2.77 (14) &1.92 (16) &2.99 (9) &2.24 (10) &0.76 (8) &10.24 (9) &3.80 (3) &\textbf{2.72 (1)} & 0.82 (9) &0.67 (8) &0.70 (8) &75.60 (7) &2.96 (12) &\hspace*{1.6 mm }9.63 (10) \\
 13 &\texttt{T13} &G &  \textbf{3.10 (1)} &\textbf{3.74 (1)} &\textbf{2.53 (1)} &1.99 (21) &1.34 (21) &0.54 (22) &\hspace*{-1 mm }-12.28 (22) &\hspace*{-1.6 mm }10.31 (22) &7.14 (22) &0.78 (17) &  0.59 (18) &  0.47 (22) &67.87 (21) &\textbf{3.69 (1)} &12.53 (13) \\
\rowcolor[HTML]{EEEEEE} 22 &\texttt{T22} &- & 1.90 (22) &1.58 (22) &1.55 (22) &1.83 (22) &1.31 (22) &0.58 (21) &\hspace*{1.5 mm }3.24 (20) &9.34 (21) &5.84 (20) &0.70 (22) &  0.51 (22) &  0.55 (19) &73.41 (12) &2.13 (22) &20.50 (22) \\

\hdashline
  1 &\texttt{T2}'$\dagger$ &D+G &2.92 (3) &3.24 (3) &2.16 (2) &3.17 (3) &2.47 (2) &0.79 (3) & 11.10 (4) &3.96 (3) &2.99 (2) &0.84 (2) &0.70 (3) &0.74 (1) & 76.06 (4) & 3.32 (2) &2.50 (1) \\
  2 &\texttt{T3}' &D+G &2.90 (5) &3.11 (4) &2.13 (3) &3.18 (1) &2.44 (3) &0.79 (2) & 11.61 (2) &4.44 (5) &3.44 (5) &0.83 (3) &0.71 (2) &0.72 (3) & 77.51 (2) & 3.27 (3) & 3.00 (2) \\
  3 &\texttt{T5}' &D+G &2.83 (8) &2.90 (7) &2.02 (4) &3.18 (2) &2.49 (1) &0.81 (1) &11.76 (1) &3.63 (1) &3.00 (3) &0.85 (1) &0.72 (1) &0.74 (2) &78.40 (1) & 2.99 (5) & 3.07 (3) \\
  4 &\texttt{T6}' &D+G &2.91 (4) &3.31 (2) &1.96 (5) &3.00 (4) &2.31 (4) &0.78 (4) & 11.21 (3) &4.14 (4) &3.57 (6) &0.83 (4) &0.70 (4) &0.71 (4) & 76.77 (3) & 3.25 (4) & 3.87 (4) \\

\bottomrule
\end{tabular}
}
\vspace{-3mm}
\end{table*}
\begin{table*}[t]
\setlength{\tabcolsep}{3pt}
\caption{Language-wise DNSMOS and CAcc scores of selected models.}
\vspace{-3.5mm}
\label{tab: languagewise_results_track1}
\centering
\resizebox{0.92\linewidth}{!}{
\begin{tabular}{ll|cc|cc|cc|cc|cc|cc}
\toprule

 \textbf{ID} &\textbf{Model} &\multicolumn{2}{c|}{\textbf{German}} &\multicolumn{2}{c|}{\textbf{English}} &\multicolumn{2}{c|}{\textbf{French}} &\multicolumn{2}{c|}{\textbf{Spanish}} &\multicolumn{2}{c|}{\textbf{Chinese}} &\multicolumn{2}{c}{\textbf{Japanese (unseen)}}  \\

   &\textbf{type} &\textbf{DNSMOS}$^\uparrow$  &\textbf{CAcc (\%)}$^\uparrow$ &\textbf{DNSMOS}$^\uparrow$  &\textbf{CAcc (\%)}$^\uparrow$ &\textbf{DNSMOS}$^\uparrow$  &\textbf{CAcc (\%)}$^\uparrow$ &\textbf{DNSMOS}$^\uparrow$  &\textbf{CAcc (\%)}$^\uparrow$ &\textbf{DNSMOS}$^\uparrow$  &\textbf{CAcc (\%)}$^\uparrow$ &\textbf{DNSMOS}$^\uparrow$  &\textbf{CAcc (\%)}$^\uparrow$   \\

\midrule

 \texttt{T1} &D &2.90 &82.0 &2.80 &79.2 &2.86 &75.8 &2.82 &86.4 &2.98 &75.3 &2.93 &75.1 \\

 \texttt{T2} &D+G &2.93 &79.4 &2.80 &73.6 &2.92 &73.7 &2.89 &83.4 &3.01 &64.3 &2.96 &71.2 \\
 
 \texttt{T3} &D+G &2.96 &79.9 &2.84 &74.4 &2.95 &74.5 &2.89 &85.9 &3.02 &71.3 &2.99 &67.9 \\
 
 \rowcolor[HTML]{EEEEEE} \texttt{T10} &D &2.88 &79.0 &2.75 &73.8 &2.81 &70.2 &2.85 &82.3 &2.92 &70.7 &2.89 &73.0 \\
 
 \texttt{T13} &G &3.10 &74.0 &3.17 &68.0 &3.14 &68.6 &2.97 &80.9 &3.16 &20.1 &3.07 &36.8 \\
 
 \rowcolor[HTML]{EEEEEE}\texttt{T22} &- &2.02 &76.8 &1.71 &70.3 &1.76 &67.3 &1.97 &80.8 &1.97 &69.3 &1.94 &74.5 \\

\bottomrule
\end{tabular}
}
\vspace{-3.5mm}
\end{table*}

In this section, we report the evaluation results of the submitted systems.
We also provide some preliminary analysis based on the brief system descriptions we collected from participants.

\subsection{Overall results}
Table~\ref{tab: results_track1and2} shows the evaluation results, where 12 out of 22 systems submitted to \texttt{Track1} (\texttt{T*}) and 4 out of 10 systems submitted to \texttt{Track2} (\texttt{S*}) are shown due to space limitations.
The full results are available on our leaderboard\footnote{\url{https://urgent-challenge.com/competitions/13}}.
D, G, and D+G in `Model type' cell denote discriminative, generative, and hybrid (e.g., discriminative model with adversarial loss or cascade of discriminative and generative models), respectively.
\texttt{T10} and \texttt{T22} are the baseline system and noisy audios, respectively. 

Focusing on the systems that outperformed the baseline (\texttt{T1}-\texttt{T9}), we observed that most were hybrid systems.
In particular, the most commonly used approach was to optimize a model with both discriminative losses and an adversarial loss (\texttt{T2}, \texttt{T4}, \texttt{T5}, \texttt{T6}, and \texttt{T9}).
These systems can leverage the strong denoising and dereverberation capabilities of well-established discriminative models while also benefitting from generative loss to tackle distortions that are particularly well handled by generative models (e.g., bandwidth limitation, packet loss, etc.).
\texttt{T7} employed a similar framework but used denoising score-matching loss.
\texttt{T3}, which showed a strong performance on all the metrics, was a cascade of of D, G, and D systems where the second model is a discrete-token-based generative model.
However, interestingly, the top system \texttt{T1} was a purely discriminative model based on a sub-band recurrent neural network (RNN).
Although the baseline system \texttt{T10} has a similar architecture (based on full- and sub-band RNNs), notably, \texttt{T1} has $\sim$102M parameters, which is much larger than that of \texttt{T10} ($\sim$8.5M).
The larger model may have been benefited since the dataset was large.

Although larger-scale training data was available in \texttt{Track2}, no systems in \texttt{Track2} outperform the best system in \texttt{Track1}, which suggests that scaling up the data does not necessarily lead to better performance in SE when the additional data contain some noisy speech.

\subsection{Language dependency of SE models}
To analyze the language dependency of SE models, we list the DNSMOS and CAcc scores for several systems by language in Table~\ref{tab: languagewise_results_track1}.
Based on the results, the discriminative models (\texttt{T1} and \texttt{T10}) appear to be relatively insensitive to language variations.
In contrast, the purely generative model based on latent diffusion and vocoding (\texttt{T13}) demonstrates a high degree of language dependency, with its performance markedly declining when applied to an unseen language, namely Japanese.
Compared with \texttt{T13}, the hybrid approaches (\texttt{T2} and \texttt{T3}) exhibit lower language dependency.
However, while \texttt{T3}, a cascade system of discriminative and purely generative models, achieves higher CAcc than \texttt{T10} in all languages except Japanese, CAcc of \texttt{T3} for Japanese data is lower than that of \texttt{T10}, indicating that there is still room for improvement in robustness on unseen languages.

Since \texttt{T13} exhibited a markedly different trend from the others, we conducted a more detailed analysis of its outputs.
Audio examples are available on our demo page\footnote{\url{https://kohei0209.github.io/urgent25-demo/}}.
We found that the model occasionally hallucinated spoken content, particularly under low-SNR conditions, consistent with the findings in~\cite{LPS}.
Notably, for Japanese inputs—which were unseen during training—the output under low-SNR conditions sometimes resembled English or other European languages that dominated the training data.
A similar phenomenon was observed in inpainted frames affected by packet loss.
These findings suggest that generative SE models may exhibit a non-negligible degree of language dependency

\subsection{Preliminary analysis on evaluation metrics}
As expected, Table~\ref{tab: languagewise_results_track1} demonstrates that DNSMOS yields high scores even when hallucinations occur.
Although only DNSMOS is reported here due to space constraints, the other two non-intrusive metrics exhibited similar trends.
These results imply that, particularly when evaluating generative models, it is essential to combine non-intrusive metrics with other metrics that can penalize hallucination, such as CAcc.

In the subjective evaluation, some generative approaches were preferred over the top system based on the discriminative model.
Note that we conducted subjective evaluations only on English data due to the time constraint of the Interspeech challenge.
Generative models are expected not to hallucinate the content heavily on English data as the majority of the training data was English.
Since P.808 is a reference-free absolute category rating test where the prompt was “How do you rate the overall quality of the following speech sample?”, it would be difficult to penalize the correctness of the spoken content and speaker consistency as long as the speech sounds natural.
Thus, the listeners may have given higher scores to generative models, which typically output speech with less audible distortions than discriminative models.
The results suggest that a new subjective listening protocol, which takes into account spoken content and/or speaker consistency, would be needed to evaluate generative models comprehensively.

\vspace{-5pt}
\section{Conclusion}
We introduced the Interspeech 2025 URGENT Challenge to investigate language dependency, universality on various distortion types, the effectiveness of using noisy data, and the data scalability of SE models.
The most popular approach was hybrid models but purely discriminative models gave the best score.
The analysis of the submissions implied that (i) generative approaches appeared to be more language-dependent than discriminative ones, (ii) more but possibly noisy data does not necessarily lead to better performance, and (iii) generative models tended to be preferred to discriminative ones in the subjective evaluation (P.808 test).
In future work, we will perform a subjective listening test on languages other than English to further analyze the language dependency of the SE systems and metrics.

\vspace{-5pt}
\section{Acknowledgement}
This work was partially supported by JSPS KAKENHI Grant Number JP24KJ2096.
The leaderboard evaluation has been supported by the PSC Bridges2 system via ACCESS allocation CIS210014, supported by National Science Foundation grants \#2138259, \#2138286, \#2138307, \#2137603, and \#2138296.
The subjective listening test was funded and executed by Technische Universität Braunschweig.

\vspace{-5pt}
\bibliographystyle{IEEEtran}
\bibliography{mybib}

\end{document}